# Continuous monitoring of mental load during virtual simulator training for laparoscopic surgery, reflects laparoscopic dexterity. A comparative study using a novel wireless device.

**Maxim Bez** [1][†], **Neta B. Maimon** [2,3][†][*], **Denis Ddobot** [4,5], **Lior Molcho** [3], **Nathan Intrator** [3,6], **Eli Kakiashvilli** [5], **and Amitai Bickel** [4,5]

[1] Medical Corps, Israel Defense Forces, Ramat Gan, Israel.

[2] The School of Psychological Sciences, Tel Aviv University, Tel Aviv, Israel.

[3] Neurosteer LTD, Herzliya, Israel.

[4] Faculty of Medicine in the Galilee, Bar-Ilan University, Safed, Israel.

[5] Department of Surgery A, Galilee Medical Center, Nahariya, Israel.

[6] Blavatnik School of Computer Science, Tel Aviv University, Tel Aviv, Israel.

[†]These authors have contributed equally to this work and share first authorship

**\* Correspondence:**
Neta B. Maimon
netacoh3@mail.tau.ac.il

## Abstract

Introduction: Cognitive Load Theory (CLT) relates to the efficiency of individuals to manipulate the limited capacity of working memory load. Repeated training generally results in individual performance increase and cognitive load decrease, as measured by both behavioral and neuroimaging methods. One of the known biomarkers for cognitive load is frontal theta band, measured by the EEG. As simulation-based training is an effective tool for acquiring practical skills, specifically to train new surgeons in a controlled and hazard-free environment, it is important to measure participants' cognitive load to decide whether they are ready to go into a real surgery. In the present study we measured performance on a surgery simulator of medical students and interns, while their brain activity was monitored by a single-channel EEG device.

Methods: Thirty-eight medical students and interns were recruited to 3 experiments examining their behavioral performance undergoing a task with the Simbionix LAP MENTOR™ simulator, while their brain activity was measured using a single-channel EEG device with novel signal processing (Aurora by Neurosteer®). On each experiment, participants performed 3 repeats of a simulator task using laparoscopic hands. The time retention between the task was different on each experiment, to examine changes in performance and cognitive load biomarkers that occur during the task or as a results of night sleep consolidation.

# Cognitive load biomarker during surgical virtual simulator

Results: The participants' behavioral performance improved with trial repetition in all 3 experiments. In Exp. 1 and Exp. 2, the theta band activity significantly decreased with better individual performance, as exhibited by some of the behavioral measurements of the simulator. The novel VC9 biomarker (previously shown to correlate with cognitive load), exhibited a significant decrease with better individual performance shown by all behavioral measurements.

Discussion: In correspondence with previous research, theta decreased with lower cognitive load and higher performance and the novel biomarker, VC9, showed higher sensitivity to load changes. Together, these measurements might be for neuroimaging assessment of cognitive load while performing simulator laparoscopic tasks. This could potentially be expanded to evaluate efficacy of different medical simulations to provide more efficient training to medical staff and to measure cognitive and mental load in real laparoscopic surgeries.

## 1      Introduction

Any acquirement of a new skill or knowledge must be passed through Working Memory (WM) before being transferred into long-term memory (LTM). According to Cognitive Load theory (CLT, Miller, 1956), WM resources, unlike LTM, are limited in their capacity for processing or holding novel information. However, when performing a complex task, the new information elements are being processed simultaneously and not iterated, so their interactions cause a much higher WM load. According to CLT, optimizing learning process may be achieved by constructing automation of schemas in which WM load is reduced. The amount of cognitive load used is a good predictor of learning process during a task performance, and continuous measurement is of particular importance (Van Merriënboer & Sweller, 2005). For instance, practicing the same task repeatedly, will cause some of the interacting information elements to be stored together in a schematic form in LTM, so that they can be extracted and manipulated in WM as a single information element (Shiffrin and Schneider 1977).

Several methods were previously described and validated as measures of WM and cognitive load. Traditionally, subjective self-rating scales were proven as reliable assessment tools across several studies (Paas et al., 2003). However, as this tool can only be recorded crudely and retrospectively, objective assessment methods with a real-time indication of WM are in demand. Using such approach, the activity can be broken down to different components that reflect different stages of complex simulations and evaluate the efficacy of each training element. Several biological methods have been reported to successfully assess WM, such as pupil size (Van Gerven et al., 2004), eye movement tracking (Van Gog & Scheiter, 2010), salivary cortisol levels (Bong et al., 2010), and functional magnetic resonance imaging (fMRI) (Van Dillen et al., 2009). Using electroencephalography (EEG), studies repeatedly show that the theta band (4-7 Hz) measured in frontal electrodes, increases with higher cognitive load and task difficulty (Antonenko et al., 2010). Multiple studies confirmed this correspondence using a verity of cognitive tasks, including the n-back (Gevins et al., 1997), operation span task (Scharniger et al., 2017), spatial and verbal WM tasks, and through simulation of real-life experience such as driving settings (Di Flumeri et al., 2018), or flight simulators (Dussault at al., 2005).

Medical simulations are a common and widespread tool for medical education. Medical simulations can emulate common scenarios in clinical practice, and through interactive interplay and hands-on teaching, improve the effectiveness and quality of teaching for healthcare professionals (Kunkler, 2006). These simulations can be particularly beneficial for surgical staff, as they allow residents to practice and perfect complex procedures, ensuring they have enough experience and practice before real patient contact (Thompson et al., 2006). As laparoscopic surgeries demand unique eye-hands coordination and are performed while the surgeon indirectly observes the intra-abdominal contents without tactile sensation ability and through a camera view, they are ideal candidates for virtual reality simulators. Indeed, these simulators have been shown to greatly improve the surgeon's operating skills and reduce operating time (Gauger et al., 2010, Gallagher et al., 2005, Yiannakopoulou et al., 2015, Larsen et al., 2009, Tucker et al., 2002).

The aim of this study was to evaluate the relationship between cognitive load and the participants' performance and skill acquisition. We aimed to track cognitive load neuro-markers using an EEG device that will enable hands mobility, while performing surgical simulator tasks. Importantly, we wanted to test medical students and interns who, on the one hand, have great motivation and motor/cognitive abilities to exceed such a laparoscopic task and on the other, have no previous hands-on experience, neither on surgery simulators nor on real-life patients. Finally, we intended to compare "online gains": improvements in performance that occur while undergoing the task, to "offline gains": the improvements in skill acquisition preceding a consolidation during night sleep between tasks trainings, without further practice (Fraser et al., 2015, Issenberg et al., 2011, Lugassy et al., 2018).

To meet these goals, we conducted 3 experiments on medical students and interns, each of which included three trials of the same short laparoscopic task administered by a surgical simulator (Simbionix LAP MENTOR™). While performing the tasks, participants' brain activity was recorded using a wearable single-channel EEG device (Aurora by Neurosteer® Inc), from which we continuously measured frontal Theta activity and a novel cognitive load biomarker: VC9 (recently validated by Maimon et al., 2021, Molcho et al., 2021, Bolton et al., 2021). Thus, our hypotheses were the following: 1. Participants' performance will improve with the surgery simulator trial repetitions 2. Cognitive load, as measured by Theta and VC9, will decrease with simulator trial repetitions 3. Theta and VC9 activity will decrease with higher individual performance (reflecting the reduced need for cognitive capabilities together with improving laparoscopic dexterity) and 4. The "offline gains" will be present following procedural memory consolidation during night sleep. Probing these hypotheses will help reveal new and objective information regarding the efficacy of simulation-based training.

## 2 Experiment 1

## 2.1 Materials and Methods

### 2.1.1 Participants

A total of 19 participants (68% females, mean age 28, age range:25-36) were enrolled to this study. All participants were healthy medical interns who completed 6 years of medical studies, never participated in laparoscopic surgery, and without prior experience using a surgical simulator. Ethical approval for this study was granted by the Galilee Medical Center institutional review board.

### 2.1.2 EEG device

The EEG signal acquisition system included a 3-electrode patch attached to the subject's forehead (Aurora by Neurosteer®., Herzliya, Israel). The medical-grade electrode patch includes dry gel for optimal signal transduction. The electrodes are located at Fp1 and Fp2 and a reference electrode at Fpz. EEG signal was amplified by a factor of a 100 and sampled at 500 Hz. Signal processing was performed in the Neurosteer cloud.

### 2.1.3 Signal processing

Full technical specifications regarding the signal analysis are provided in the Supplementary Material. In brief, the signal processing algorithm interprets the EEG data using a time/frequency wavelet-packet analysis, instead of the commonly used spectral analysis. It is computationally involved and relies on a previously collected large cohort of EEG data from which the features were extracted. 121 Brain Activity Features (BAFs) were created using a variant of the wavelet packet analysis and the best basis algorithm (Coifman & Wickerhauser, 1992). The creation of the biomarkers is based on another large cohort of independent data from healthy subjects performing different tasks, collected with the same device. The labelled data was used in a supervised machine learning procedure to identify biomarkers of equal and nonequal-weight BAFs. Out of these different biomarkers, VC9 was recently validated as a cognitive load biomarker using the n-back task, auditory detection task, and interruption task (Maimon et al., 2021, Molcho et al., 2021 and Bolton et al., 2021). It was shown that VC9 activity increased with increasing levels of cognitive load within cognitively healthy participants and showed higher sensitivity than the theta band. in this study we used VC9 as well as frontal theta activity (amplitude in 4-7 Hz) measured by the device.

### 2.1.4 Procedure

The Aurora Electrode strip with 3 frontal electrodes was attached to the subject's forehead and connected to the device for brain activity recording. The participants were then asked to complete a task with a surgical simulator Simbionix LAP MENTOR™ (Simbionix, Airport City, Israel) which involved grasping and clamping of blood vessels using two different laparoscopic arms. The same task was used for all subjects. At the end of the task, the participants were rated by the surgical simulator based on three main parameters: accuracy, economy of movement, and time required to complete the task. All participants were monitored by the EEG device and were given 3 attempts to perform the same task. Participants performed three consecutive trials on the same session, with a five-minute break between the trials. The performance of each participant was scored by the surgical simulator's algorithm based on their accuracy, economy of movement, and time to complete the task. The concurrent activity of VC9 biomarker and theta band was extracted by the Aurora system.

### 2.1.5 Statistical Analysis

Behavioral performance was extracted from the surgical simulator after each trial of each participant, including accuracy (in percentages), economy of movement (in percentages) and time to exceed the trial (in seconds). Theta (4-7 Hz) and VC9 activity were extracted from the Aurora system. Activity of all the dependent variables was averaged per each trial per each participant and reported as mean and standard deviation. Two analyses were performed on the behavioral and EEG data. Six mixed linear models (Boisgontier & Cheval, 2016) were designed (one on each dependent variable) to measure the differences between the three simulator trials for each dependent variable. The model used trial number (1,2, or 3) as a within-participants variable, and participants were inserted into the random slope. Since trial number variable included 3 levels, indicator variables (aka dummy variables) were computed with trial 1 as the reference group. Accordingly, two effects resulted from the model: one of the 2[nd] trial which represents the significance of difference between the first and second trials, and one of the 3[rd] trial which represents the significance of difference between the first and third trials. Next, Pearson's correlation coefficients were calculated to evaluate the correlation between each of the EEG variables activity and behavioral performance of each participant per each trial. Two-tailed p<0.05 was considered statistically significant. Analyses were performed using Python Statsmodels (Seabold & Perktold, 2010).

### 2.2 Results

### 2.2.1 Behavioral measurements

A full description of the Mixed Linear Models (LMM) parameters (Coefficients, standard errors, Z scores, *p* values and confidence intervals) for all models performed in this study are presented in table 1.

A significant increase of the participants' accuracy was observed between the trials (p<0.001; Figure 1), as well as a significant increase in the economy of movement (p<0.001). The average time required for the completion of the task was also significantly reduced between the trials (p<0.001).

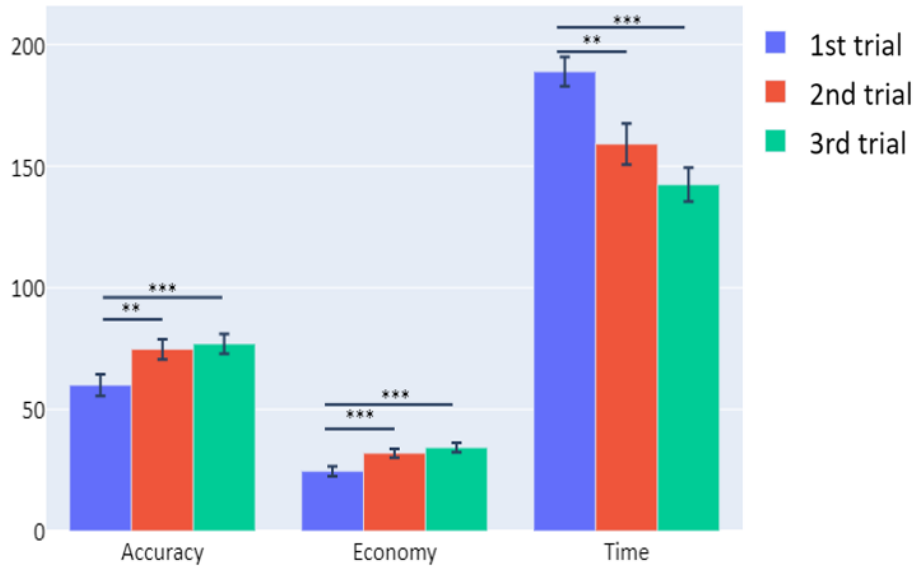

**Figure 1:** The mean and SE of accuracy (%) , economy measures (%), and time (sec) in Exp. 1, for trial 1 (blue), trial 2 (red) and trial 3 (green), in the first session for all participants (n=19).

### 2.2.2 EEG features

VC9 activity was significantly reduced between the first and the third attempts (*p*=0.011 and *p*=0.021 for the differences between first and second trials, and first and third trials respectively, see figure 2). Theta did not exhibit any significant effect between the trails (all *p*s>0.05).

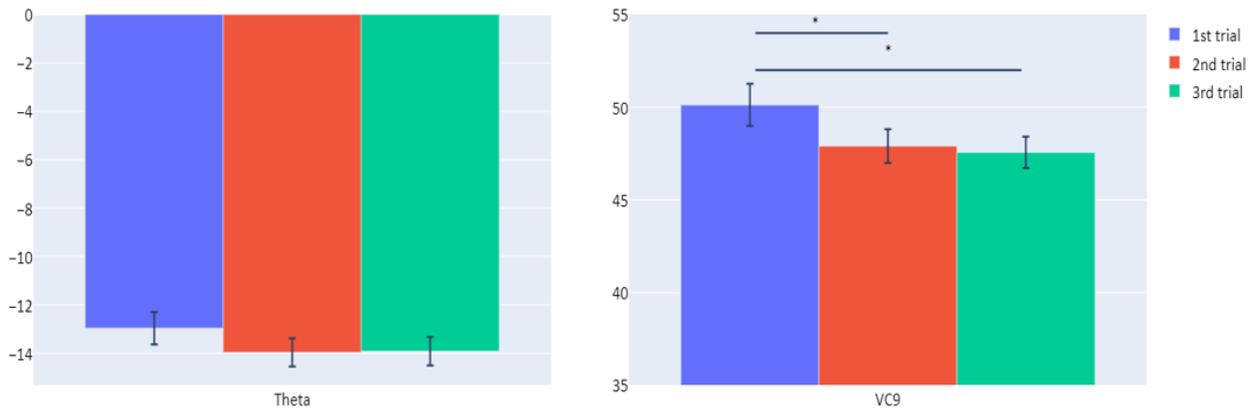

**Figure 2**: The mean and SE of VC9 and Theta, EEG features in Exp. 1 for trial 1 (blue), trial 2 (red) and trial 3 (green) in experiment 1.

### 2.2.3 Correlations between EEG features activity and Behavioral performance

VC9 activity significantly correlated with all three behavioral measurements and was found to decrease with better performance, i.e., negatively correlated with accuracy and economy of movement and positively correlated with time: r = -0.57, $p<0.001$, r = -0.56, $p<0.001$ and r = 0.297, $p=0.025$ for accuracy, economy of movement and time respectively; see Figure 3. Theta decreased with higher accuracy and economy of movement, r = -0.496, $p<0.001$, r = -0.498, $p<0.001$, for accuracy and economy of movement, respectively. However, it did not increase with longer time: r = 0.25, $p=0.058$.

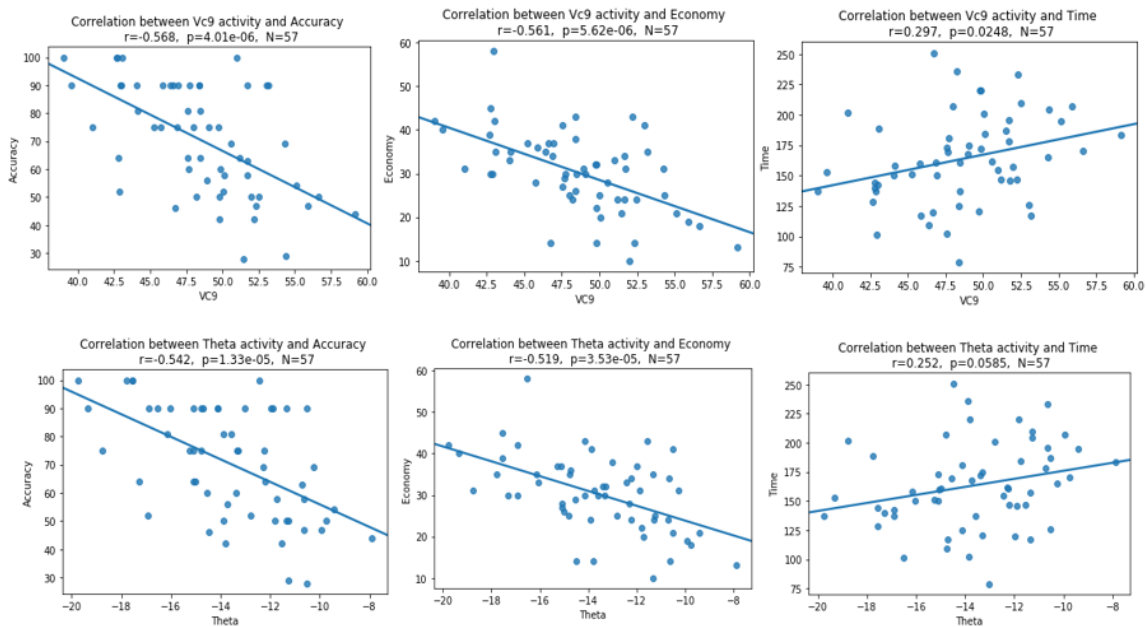

**Figure 3:** Mean activity of the VC9 (top) and Theta (bottom) in Exp. 1, as function of Accuracy (A), Economy (B) and Time (C). Pearson R and $p$ values presented.

## 2.3 Discussion

Exp. 1 results exhibited a significant improvement in all behavioral measurement of performance, as well as a significant decrease VC9 activity (but not Theta) between the three trials. In addition, VC9 activity decreased with better individual performance, which was expressed by the significant correlations between all three behavioral measures and VC9 activity. Theta also exhibited individual differences and decreased with higher accuracy and economy of movement and marginally increased with longer time to complete the task. These results suggest that changes in cognitive load are corresponding to performance in a surgery simulator as shown by VC9 and, to some extent, frontal theta.

Next, we aimed to explore the effect of night sleep memory consolidation on task performance in the simulator and participants' brain activity. Hence, we conducted a second experiment which included ten participants who also participated in Exp. 1. They performed an additional three-trial session on the consecutive day.

## 3 Experiment 2

### 3.1 Materials and Methods

#### 3.1.1 Participants and procedure

On the consecutive day of Exp. 1, ten randomly chosen participants performed an additional 3-trials session with the same procedure.

### 3.2 Results

#### 3.2.1 Behavioural Measurements

Behavioral performance did not differ between the three trials of this session in any of the measurements (all ps >0.05, see Figure 4).

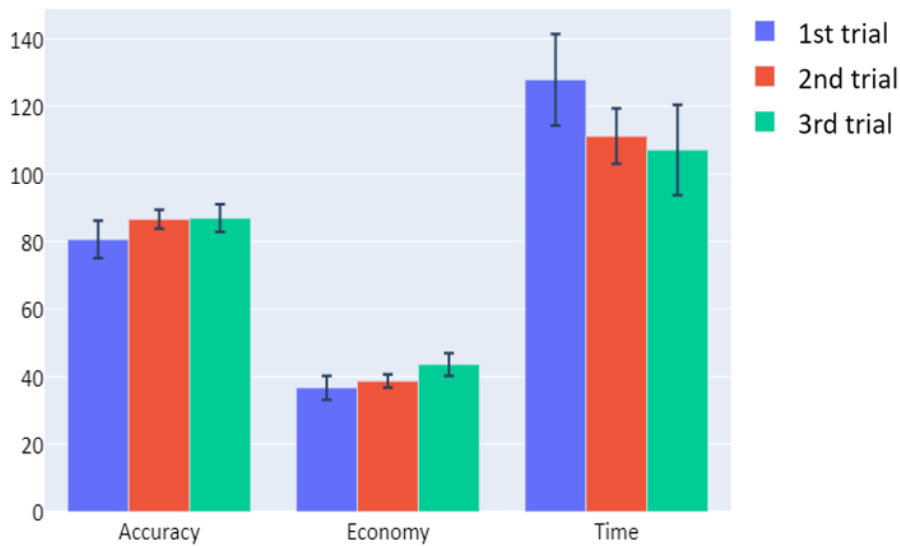

**Figure 4:** The mean and SE of accuracy (%), economy measures (%) and time (sec) in Exp. 2, for trial 1 (blue), trial 2 (red) and trial 3 (green), in the first session for all participants (n=10).

### 3.2.2 EEG features

Activity levels of all EEG features (VC9 and Theta) did not differ between the trials of the session, *p*>0.05, (see Figure 5).

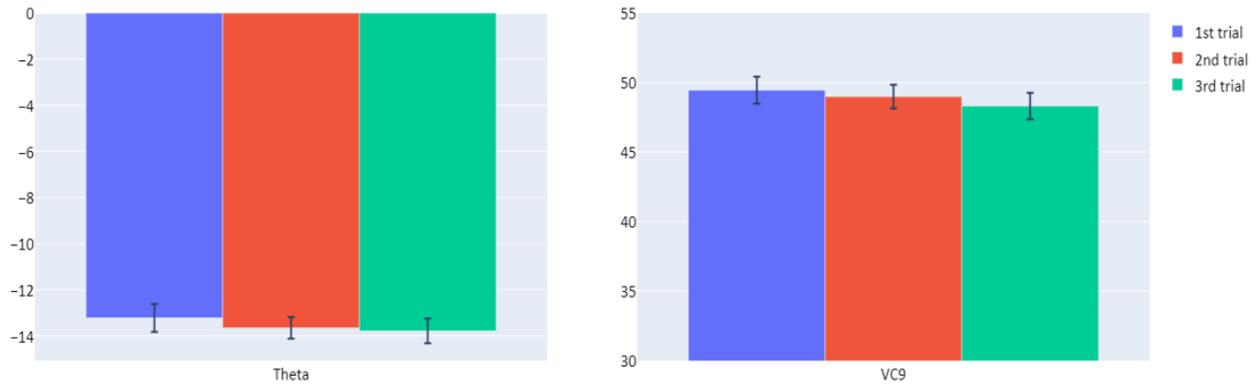

**Figure 5**: Means and SE of EEG features VC9 and Theta in Exp. 2 for trial 1 (blue), trial 2 (red) and trial 3 (green).

### 3.2.3 Correlations between EEG features activity and Behavioral performance

VC9 activity significantly correlated with all three behavioral measurements and was found to decrease with better participants performance, i.e. negatively correlated with accuracy and economy of movement and positively correlated with time: r = -0.575, p=0.001, r = -0.42, p=0.022 and r = 0.4, *p*=0.027 for accuracy, economy of movement and time respectively; Theta significantly reduced with higher accuracy (r=-0.525, *p*=0.003) but did not correlate significantly with either economy of movement or time (both ps>0.05, see Figure 6).

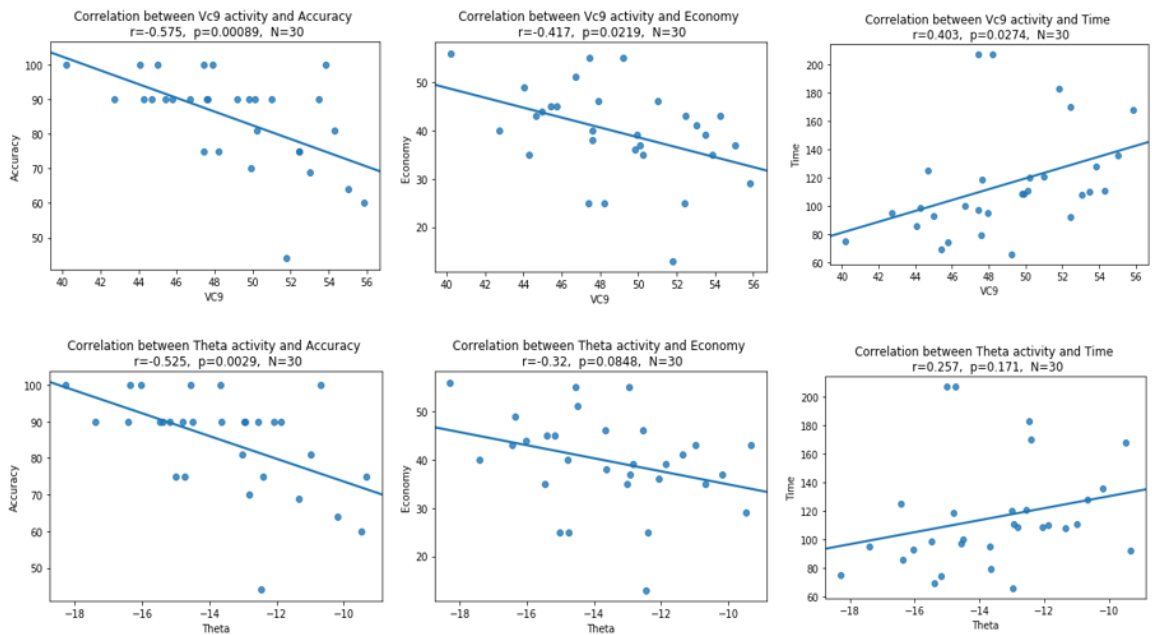

**Figure 6:** Mean activity of the VC9 (top) and Theta (bottom) in Exp. 2, as function of Accuracy (A), Economy (B) and Time (C). Pearson R and *p* values presented.

### 3.2.4 Comparison between Exp 1 and Exp 2

To explore the "offline gains" between Exp. 1 and Exp 2, behavioral performance and brain activity of the 10 participants who participated in both experiments was compared. Four paired t-tests comparing accuracy, economy of movement, time and VC9 activity for the last trial of Exp 2 and the first trial of Exp. 3 were conducted. Both behavioral performance and brain activity were the same in the first trial of Exp 2 relatively to the last trial of Exp 1: all ps >0.05, see figure 7.

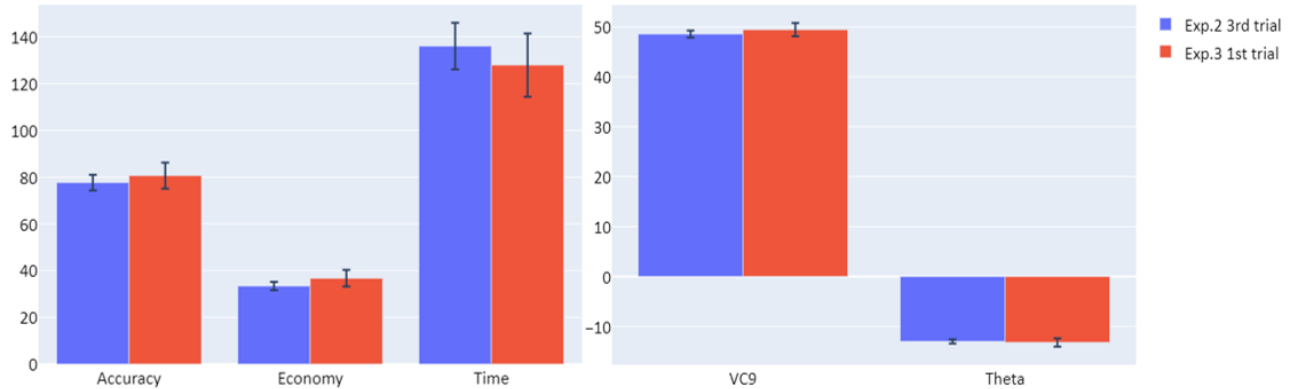

**Figure 7:** The means and SE of accuracy, time, economy (left), and VC9 and Theta (right), in the third trial of Exp 1 (blue) and first trial of Exp 2 (red), averaged over the 10 participants who underwent both experiment (n=10).

### 3.3 Discussion

Exp. 2 results exhibited a clear correspondence between behavioral performance and both EEG features. First, both behavioral performance and VC9 activity, did not exhibit significant differences between the three trials, meaning that participants performance did not improve, and theta and VC9 activity did not decrease between the three trials. Second, VC9 activity decreased with better behavioral performance as depicted by significant correlation with all three behavioral measurements. Theta correlated with accuracy but not with economy of movement and time.

The comparison between the last trial of Exp.1 and first trial of Exp. 2 showed no difference between the two in any of the behavioral or neurophysiological measures. However, taken together with the null improvement during these three sessions, this lack of difference may be explained with participants' reaching their asymptotic level (ceiling effect): participants reached their maximum ability on the third trial of the first day, and maintained it in the first to last trials of the second day. To further reveal "offline gains" from night sleep consolidation, we conducted a third experiment. New participants performed a single task trial on three consecutive days. This was done to make sure participants will not reach asymptotic performance levels before the night sleep consolidation.

## 4 Experiment 3

### 4.1 Materials and Methods

### 4.1.1 Participants and procedure

19 (63% females) healthy medical students from first to sixth year of studying (mean=4), with mean age of 25.631 (SD=2.532). All students had no prior experience using a surgical simulator. Participants underwent the same task as Exps. 1 & 2, performing one trial per day for three consecutive days.

## 4.2 Results

### 4.2.1 Behavioral measurements

A significant increase in participants' accuracy results was observed between the attempts (p<0.001; Figure 8), as well as a significant increase in the economy of movement (p<0.001). The average time required for the completion of the task was also significantly reduced between the attempts (p<0.001).

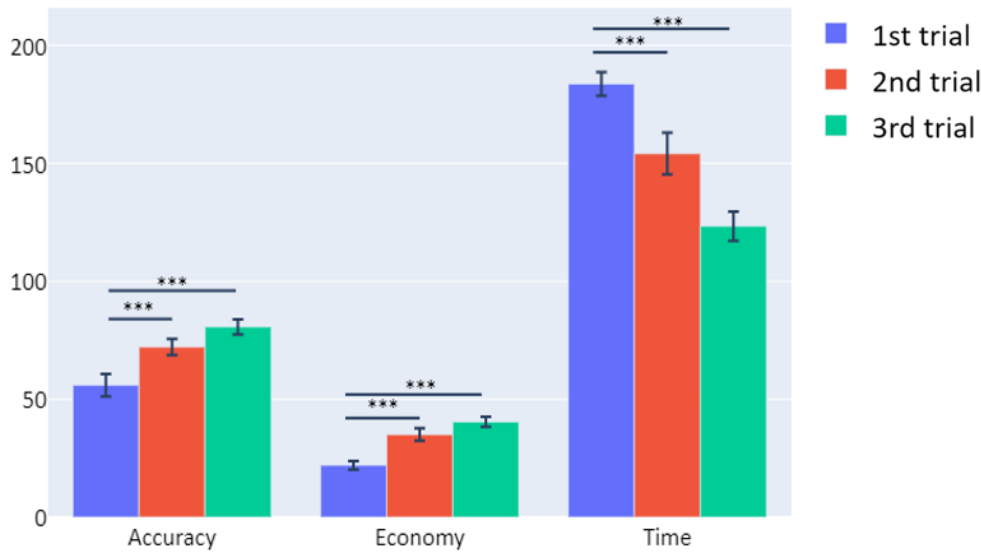

**Figure 8**: The mean and SE of accuracy (%), economy (%), and time (sec) measures in experiment 3 for trial 1 (day 1, blue), trial 2 (day 2, red) and trial 3 (day 3, green).

### 4.2.2 EEG features

There was no significant difference in VC9 activity between the trials in consecutive days, although the difference between the first and second attempt was marginally significant ($p=0.056$ and $p>0.05$ for the differences between the first and second trials, and the first and third trials, respectively, see figure 9). Theta exhibited a significant difference between the first and second trials ($p=0.019$), but the difference between the first and third trials did not reach significance level ($p>0.05$).

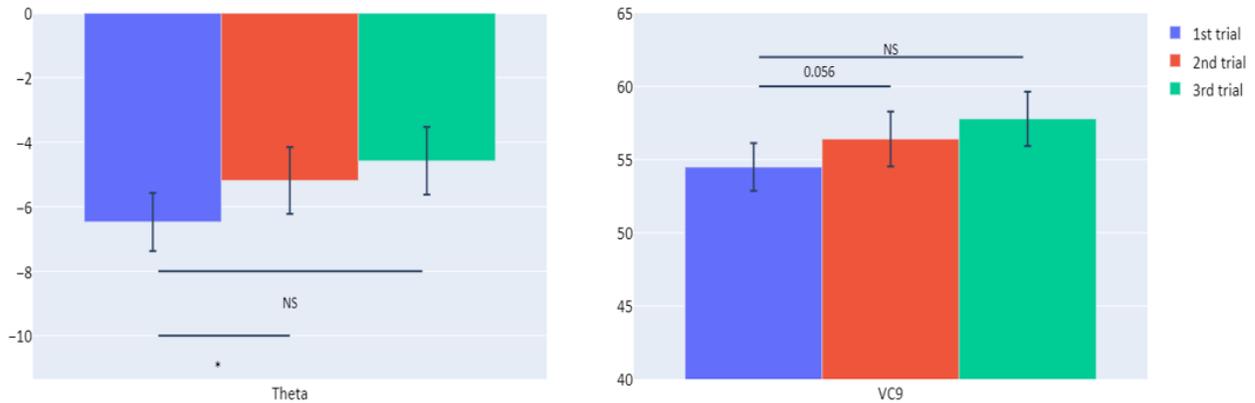

**Figure 9**: The mean and SE of EEG features VC9 and Theta in Exp. 3 for trial 1 (blue), trial 2 (red) and trial 3 (green).

### 4.2.3 Correlations between EEG features activity and Behavioral performance

No significant correlations were found between any of the EEG features and behavioral measurement (VC9: r = -.19, r = -0.19 and r = 0.1 for accuracy, economy of movement, and time respectively, Theta: r = -.19, r = -0.2 and r = 0.09 for accuracy, economy of movement, and time respectively, all $p$s>0.05).

| Experiment | Feature | Effect | Coef. | Std.Err. | z | P>|z| | [0.025 | 0.975] |
|---|---|---|---|---|---|---|---|---|
| *Exp. 1* | Accuracy | Intercept | 61.837 | 4.143 | 14.926 | <0.001 | 53.717 | 69.957 |
| | | 1st trial vs. 2nd trial | 12.873 | 3.838 | 3.354 | 0.001 | 5.351 | 20.396 |
| | | 1st trial vs. 3rd trial | 17.473 | 3.422 | 5.106 | <0.001 | 10.765 | 24.181 |
| | Economy | Intercept | 25.18 | 1.931 | 13.039 | <0.001 | 21.395 | 28.965 |
| | | 1st trial vs. 2nd trial | 6.82 | 1.892 | 3.605 | <0.001 | 3.112 | 10.528 |
| | | 1st trial vs. 3rd trial | 7.968 | 1.894 | 4.206 | <0.001 | 4.255 | 11.681 |
| | Time | Intercept | 185.841 | 5.33 | 34.867 | <0.001 | 175.395 | 196.288 |
| | | 1st trial vs. 2nd trial | -26.789 | 9.513 | -2.816 | 0.005 | -45.434 | -8.144 |
| | | 1st trial vs. 3rd trial | -40.998 | 8.638 | -4.746 | <0.001 | -57.929 | -24.068 |
| | VC9 | Intercept | 49.531 | 1.006 | 49.215 | <0.001 | 47.558 | 51.503 |
| | | 1st trial vs. 2nd trial | -1.506 | 0.594 | -2.537 | 0.011 | -2.67 | -0.343 |
| | | 1st trial vs. 3rd trial | -1.75 | 0.757 | -2.313 | 0.021 | -3.233 | -0.267 |
| | Theta | Intercept | -13.324 | 0.625 | -21.314 | <0.001 | -14.549 | -12.099 |
| | | 1st trial vs. 2nd trial | -0.556 | 0.407 | -1.367 | 0.172 | -1.354 | 0.241 |
| | | 1st trial vs. 3rd trial | -0.397 | 0.589 | -0.675 | 0.5 | -1.551 | 0.756 |
| *Exp. 2* | Accuracy | Intercept | 84.871 | 4.185 | 20.279 | <0.001 | 76.668 | 93.074 |
| | | 1st trial vs. 2nd trial | 1.729 | 4.531 | 0.382 | 0.703 | -7.151 | 10.609 |
| | | 1st trial vs. 3rd trial | 2.059 | 3.817 | 0.539 | 0.59 | -5.422 | 9.539 |
| | Economy | Intercept | 39.225 | 2.637 | 14.873 | <0.001 | 34.056 | 44.394 |
| | | 1st trial vs. 2nd trial | -0.525 | 3.49 | -0.15 | 0.88 | -7.365 | 6.316 |
| | | 1st trial vs. 3rd trial | 4.375 | 3.412 | 1.282 | 0.2 | -2.312 | 11.062 |
| | Time | Intercept | 119.596 | 11.414 | 10.478 | <0.001 | 97.224 | 141.968 |
| | | 1st trial vs. 2nd trial | -8.396 | 12.449 | -0.674 | 0.5 | -32.796 | 16.004 |
| | | 1st trial vs. 3rd trial | -22.883 | 12.971 | -1.764 | 0.078 | -48.306 | 2.54 |
| | VC9 | Intercept | 49.451 | 1.341 | 36.874 | <0.001 | 46.823 | 52.079 |
| | | 1st trial vs. 2nd trial | -0.466 | 0.726 | -0.642 | 0.521 | -1.889 | 0.957 |
| | | 1st trial vs. 3rd trial | -0.758 | 0.966 | -0.784 | 0.433 | -2.652 | 1.136 |
| | Theta | Intercept | -13.218 | 0.824 | -16.043 | <0.001 | -14.833 | -11.603 |
| | | 1st trial vs. 2nd trial | -0.433 | 0.611 | -0.709 | 0.478 | -1.63 | 0.764 |
| | | 1st trial vs. 3rd trial | -0.357 | 0.638 | -0.56 | 0.576 | -1.608 | 0.894 |
| *Exp. 3* | Accuracy | Intercept | 59.067 | 4.008 | 14.738 | <0.001 | 51.211 | 66.922 |
| | | 1st trial vs. 2nd trial | 13.023 | 3.715 | 3.505 | <0.001 | 5.741 | 20.305 |
| | | 1st trial vs. 3rd trial | 21.607 | 3.887 | 5.559 | <0.001 | 13.99 | 29.225 |
| | Economy | Intercept | 22.003 | 1.845 | 11.923 | <0.001 | 18.386 | 25.62 |
| | | 1st trial vs. 2nd trial | 14.209 | 1.677 | 8.47 | <0.001 | 10.921 | 17.497 |
| | | 1st trial vs. 3rd trial | 17.665 | 2.305 | 7.664 | <0.001 | 13.147 | 22.183 |
| | Time | Intercept | 182.022 | 4.539 | 40.103 | <0.001 | 173.127 | 190.918 |
| | | 1st trial vs. 2nd trial | -31.743 | 7.613 | -4.17 | <0.001 | -46.664 | -16.822 |
| | | 1st trial vs. 3rd trial | -58.707 | 5.368 | -10.936 | <0.001 | -69.228 | -48.185 |
| | VC9 | Intercept | 54.489 | 1.633 | 33.375 | <0.001 | 51.289 | 57.689 |
| | | 1st trial vs. 2nd trial | 1.915 | 1.002 | 1.911 | 0.056 | -0.049 | 3.88 |
| | | 1st trial vs. 3rd trial | 2.493 | 1.967 | 1.267 | 0.205 | -1.362 | 6.348 |
| | Theta | Intercept | -6.473 | 0.899 | -7.197 | <0.001 | -8.235 | -4.71 |
| | | 1st trial vs. 2nd trial | 1.285 | 0.548 | 2.346 | 0.019 | 0.211 | 2.358 |
| | | 1st trial vs. 3rd trial | 1.432 | 1.034 | 1.385 | 0.166 | -0.594 | 3.459 |

**Table 1**: Coefficients, standard errors, Z scores, *p* values and confidence intervals for all effects depicted in the three LMM analyses applied in the 3 experiments of the present study. All LMMs included trial (1/2/3) as a within-participants variable, with encoded effects of trial 1 vs. trial 2 and trial 1 vs. trial 3.

### 4.3 Discussion

Exp. 3 results reflect "offline gains": participants' performance under the simulator improved with trials repetition, although each occurred on the consecutive day. However, VC9 and Theta did not show a significant difference between the trials or exhibit significant correlations with any of the behavioral measurements. This discrepancy might be explained by the different brain networks that take part in "online" versus "offline" learning, see general discussion below.

## 5 General Discussion

In this study, we continuously measured cognitive load levels using theta band power and VC9 biomarker, during task performance on a surgical simulator. A single-channel EEG device utilizing decomposition of the EEG signal via Harmonic Analysis was used to obtain the novel VC9 biomarker as well as the frequently used theta band. Both EEG features, theta and VC9, were previously validated as objective measures for cognitive load levels (Antonenko et al., 2010, Maimon et al., 2021, Molcho et al., 2021, Bolton et al., 2021). Exp 1 consisted of 3 repeating trials on the surgery simulator, and results show participants' performance improved and VC9 activity decreased. Additionally, VC9 and, to some extent, theta band, decreased with higher individual performance. Overall, these results support previous finding that as the participant becomes more proficient in performing the task, depicted by the behavioral performance, the pre-frontal activity associated with cognitive load is reduced (Takeuchi et al., 2013). To examine the effect of night sleep memory consolidation, we conducted Exp. 2, which included 3 simulator trials on the consecutive day. Results revealed no difference in performance or EEG features between the three trials. Still, VC9 and, to some extent, theta band, exhibited significant correlations to individual performance, even without the difference between the three trials. Further analysis revealed that both behavioral and EEG features results remained similar between the last trial of the first day and the first trial on the second day . Testing the "offline gains" refers to the improvements in skill acquisition preceding a consolidation during night sleep between tasks trainings, without further practice. Taken together, these findings potentially indicate that 'offline gains' were not detected since participants reached their maximum performance level on the last trial of the first day and maintained it throughout the trials in the following day.

Therefore we conducted an additional experiment in which 19 medical students performed a single simulator trial on 3 consecutive days. This was done to prevent participants from reaching their maximum performance level on the first day, and to reveal the night sleep memory consolidation related to 'offline gains'. Indeed, participants' performance in all behavioral measurements improved significantly between the testing days. Interestingly, VC9 showed no difference between the trials and the theta band even showed a mild increase in activity between the trials. This lack of increase in activity can be explained by the difference in brain networks that are involved in fast vs. slow stages of motor skill learning (Dayan & Cohen, 2011). Similar to online learning, fast motor skills acquisition occurs during task training and could last minutes (Karni et al., 1995). On the slow stage, further gains are achieved across multiple sessions of training, mostly divided by night sleep consolidation. The neural substrates during the fast stage shows complex brain activation pattern. First, it includes an increasing activity in the supplementary motor area (SMA), dorsomedial striatum (DMS), premotor cortex (PM), posterior parietal cortex (PPC), and posterior cerebellum. This reflects the requirement of additional cortical brain activity during practice. At the same time, the fast stage also creates decreased activity of the dorsolateral prefrontal cortex (DLPFC), primary motor cortex (M1), and presupplementary motor area (preSMA). These decreases may suggest that with online practice, one uses fewer neuronal resources (Hikosaka et al., 2002). Conversely, the slow motor skill stage is characterized with increased activity in M1, S1, SMA, and DLS, and decreased activity in the lateral cerebellum. Notably, the decrease in the frontal area (DLPFC), which commonly correlate with WM load (Manoach et al., 1997), is not visible during this slow leaning stage. Thus, progress from fast to slow stages can be generalized to the shift from anterior to more posterior brain regions (Floyer-Lea and Matthews, 2005). The current WM biomarkers which were used in the present research, were extracted via frontal single-EEG channel, and therefore reflect frontal activity associated with WM load. Taken together, this may explain the results obtained in Exp. 3 of the present study, which revealed prominent 'offline gains' within the behavioral performance of the task, but no decrease within the current EEG WM load biomarkers. Further research, however, should look for novel biomarkers that will be able to correlate with such posterior/limbic activity.

The activation patterns obtained in Exp. 1 and 2 correlate with WM load as previously described (Zarjam et al., 2011 and Wang et al., 2016). Theta and VC9 activity decreased with lower cognitive load that resulted from simulator trial repetition. As shown in previous studies (Maimon et al., 2021 and Molcho et al., 2021), VC9 was found to be a more sensitive biomarker than theta as it was significantly decreased with trials repetition of Exp. 1 and significantly correlated with all behavioral measurements depicting individual performance in Exp. 1 and 2. This further validates VC9 as an effective biological measurement for the assessment of cognitive load while performing laparoscopic tasks using the surgical simulator specifically, and suggest that it may provide a measure of cognitive functioning during surgery. Due to the non-obtrusive nature of mobile EEG devices, such devices can be used by surgeons during live operations. Intraoperative evaluation can provide an objective metric to assess surgeons' performance in real life scenarios, and measure additional unobservable behaviors, such as mental readiness (Cha & Yu, 2021).

Seeing as simulation-based training is an effective tool for acquiring practical skills, the question remains as to which methods should be utilized to optimize this process and achieve better assessment of improved manual dexterity. Specifically, it is not fully understood which brain processes during simulation-based training translate to better skill acquisition through practice. Here we show that VC9 can be reliably used to monitor and assess participants' WM levels during manual practice. Furthermore, the correlations between VC9 and improved simulator scores can be translated to other areas and used on other procedures which require manual training. Interestingly, the lower baseline activity of this marker during the first trial between medical students and medical interns, which have already participated in surgeries and performed real procedures on patients, can be further explored and potentially be used as a predictor for the participant's skill in certain manual procedures.

This study has several limitations. The analyses were performed on young medical internists, which may not accurately reflect the overall medical population. Additionally, evaluating finer differences between the experiments require larger cohort of participants. Future studies on a larger and more diverse population should further validate the findings presented in this work as well as study the effect of prolonged breaks on the activity of these biomarkers. Moreover, a 24-hour break period between the sessions may not be sufficient to study the effect of long-term memory on the prefrontal brain activity during the simulation. Future studies should evaluate a longer time windows to assess the temporal changes in the activity of the working memory associated biomarkers and search for additional biomarkers to represent other cerebral regions beyond the frontal regions (like the limbic or motor system).

To conclude, in this study we extracted the previously validated cognitive load biomarkers, theta band and VC9, from a novel single-channel EEG using advanced signal analysis. Results showed high correlations of the biomarkers with participants' individual performance using a surgical simulator. As surgical simulations allow doctors to gain important skills and experience needed to perform procedures without any patient risk, evaluation and optimization of these effects on medical staff are crucial. This could potentially be expanded to evaluate efficacy of different medical simulations to provide more efficient training to medical staff and to measure cognitive and mental loads in real laparoscopic surgeries.

## 6 Conflict of Interest

Maxim Bez: no discloser, Neta B. Maimon who designed and conducted the analyses in this study is a part time researcher at Neurosteer LTD which supplied the EEG system. Denis Ddobot: no discloser, Lior Molcho: clinical researcher at Neurosteer LTD, Nathan Intrator: founder on Neurosteer LTD, Eli Kakiashvilli: no discloser, and Amitai Bickel: no discloser.

## 7 Author Contributions

A.B., M.B., N.M. and N.I. conceived and planned the experiments. A.B., M.B., and D.D. carried out the experiments. N.I., and L.M. designed and provided tech support with the EEG device. M.B., N.M., and L.M. contributed to the interpretation of the results. M.B. and N.M. took the lead in writing the manuscript. All authors provided critical feedback and helped shape the research, analysis and manuscript.

## 8 References

1. Antonenko, P., Paas, F., Grabner, R., & Van Gog, T. (2010). Using electroencephalography to measure cognitive load. Educational Psychology Review, 22(4), 425-438.
2. Boisgontier, M. P., & Cheval, B. (2016). The ANOVA to mixed model transition. Neuroscience & Biobehavioral Reviews, 68, 1004-1005.
3. Bolton, F., Te'Eni, D., Maimon, N. B. & Toch, E. (2021, March). Detecting interruption events using EEG. In 2020 IEEE 2nd Global Conference on Life Sciences and Technologies (LifeTech). IEEE.
4. Bong, C. L., Lightdale, J. R., Fredette, M. E., & Weinstock, P. (2010). Effects of simulation versus traditional tutorial-based training on physiologic stress levels among clinicians: a pilot study. Simulation in Healthcare, 5(5), 272-278.
5. Cha, J. S., & Yu, D. (2021). Objective Measures of Surgeon Nontechnical Skills in Surgery: A Scoping Review. Human Factors, 0018720821995319.
6. Coifman, R. R., & Wickerhauser, M. V. (1992). Entropy-based algorithms for best basis selection. IEEE Transactions on information theory, 38(2), 713-718.
7. Constantinidis, C., & Klingberg, T. (2016). The neuroscience of working memory capacity and training. Nature Reviews Neuroscience, 17(7), 438.
8. Dayan, E., & Cohen, L. G. (2011). Neuroplasticity subserving motor skill learning. Neuron, 72(3), 443-454.
9. Di Flumeri, G., Borghini, G., Aricò, P., Sciaraffa, N., Lanzi, P., Pozzi, S., ... & Babiloni, F. (2018). EEG-based mental workload neurometric to evaluate the impact of different traffic and road conditions in real driving settings. Frontiers in human neuroscience, 12, 509.
10. Dussault, C., Jouanin, J. C., Philippe, M., & Guezennec, C. Y. (2005). EEG and ECG changes during simulator operation reflect mental workload and vigilance. Aviation, space, and environmental medicine, 76(4), 344-351.
11. Floyer-Lea, A., and Matthews, P.M. (2005). Distinguishable brain activation networks for short- and long-term motor skill learning. J. Neurophysiol. 94, 512–518.
12. Fraser KL, Ayres P, Sweller J. Cognitive load theory for the design of medical simulations. Simulation in Healthcare. 2015.

Cognitive load biomarker during surgical virtual simulator13. Gallagher, A. G., Ritter, E. M., Champion, H., Higgins, G., Fried, M. P., Moses, G., ... & Satava, R. M. (2005). Virtual reality simulation for the operating room: proficiency-based training as a paradigm shift in surgical skills training. Annals of surgery, 241(2), 364.
14. Gauger, P. G., Hauge, L. S., Andreatta, P. B., Hamstra, S. J., Hillard, M. L., Arble, E. P., ... & Minter, R. M. (2010). Laparoscopic simulation training with proficiency targets improves practice and performance of novice surgeons. The American journal of surgery, 199(1), 72-80.
15. Gevins, A., Smith, M. E., McEvoy, L., & Yu, D. (1997). High-resolution EEG mapping of cortical activation related to working memory: effects of task difficulty, type of processing, and practice. Cerebral cortex (New York, NY: 1991), 7(4), 374-385.
16. Gurusamy, K. S., Aggarwal, R., Palanivelu, L., & Davidson, B. R. (2009). Virtual reality training for surgical trainees in laparoscopic surgery. Cochrane database of systematic reviews, (1).
17. Herff, C., Heger, D., Fortmann, O., Hennrich, J., Putze, F., & Schultz, T. (2014). Mental workload during n-back task—quantified in the prefrontal cortex using fNIRS. Frontiers in human neuroscience, 7, 935.
18. Hikosaka, O., Rand, M.K., Nakamura, K., Miyachi, S., Kitaguchi, K., Sakai, K., Lu, X., and Shimo, Y. (2002). Long-term retention of motor skill in macaque
19. Issenberg, S. B., Ringsted, C., Østergaard, D., & Dieckmann, P. (2011). Setting a research agenda for simulation-based healthcare education: a synthesis of the outcome from an Utstein style meeting. Simulation in Healthcare, 6(3), 155-167.
20. Jaeggi, S. M., Buschkuehl, M., Perrig, W. J., & Meier, B. (2010). The concurrent validity of the N-back task as a working memory measure. Memory, 18(4), 394-412.
21. Jolles, D. D., van Buchem, M. A., Crone, E. A., & Rombouts, S. A. (2013). Functional brain connectivity at rest changes after working memory training. Human brain mapping, 34(2), 396-406.
22. Karni, A., Meyer, G., Jezzard, P., Adams, M.M., Turner, R., and Ungerleider, L.G. (1995). Functional MRI evidence for adult motor cortex plasticity during motor skill learning. Nature 377, 155–158.
23. Kunkler, K. (2006). The role of medical simulation: an overview. The International Journal of Medical Robotics and Computer Assisted Surgery, 2(3), 203-210.
24. Larsen, C. R., Soerensen, J. L., Grantcharov, T. P., Dalsgaard, T., Schouenborg, L., Ottosen, C., ... & Ottesen, B. S. (2009). Effect of virtual reality training on laparoscopic surgery: randomised controlled trial. Bmj, 338.
25. Lugassy, D., Herszage, J., Pilo, R., Brosh, T., & Censor, N. (2018). Consolidation of complex motor skill learning: evidence for a delayed offline process. Sleep, 41(9), zsy123.
26. Maimon, N. B., Molcho, L., Intrator, N., & Lamy, D. (2020). Single-channel EEG features during n-back task correlate with working memory load. arXiv preprint arXiv:2008.04987.
27. Manoach, D. S., Schlaug, G., Siewert, B., Darby, D. G., Bly, B. M., Benfield, A., ... & Warach, S. (1997). Prefrontal cortex fMRI signal changes are correlated with working memory load. Neuroreport, 8(2), 545-549.
28. Molcho, L., Maimon, N. B., Pressburger, N., Regev-Plotnik, N., Rabinowicz, S., Intrator, N., & Sasson, A. (2020). Detection of cognitive decline using a single-channel EEG system with an interactive assessment tool. preprint, Research Square.
29. monkeys and humans. Exp. Brain Res. 147, 494–504.
30. Paas, F., Tuovinen, J. E., Tabbers, H., & Van Gerven, P. W. (2003). Cognitive load measurement as a means to advance cognitive load theory. Educational psychologist, 38(1), 63-71.
16This is a provisional file, not the final typeset article

31. Scharinger, C., Soutschek, A., Schubert, T., & Gerjets, P. (2017). Comparison of the working memory load in n-back and working memory span tasks by means of EEG frequency band power and P300 amplitude. Frontiers in human neuroscience, 11, 6.
32. Scheeringa, R., Petersson, K. M., Oostenveld, R., Norris, D. G., Hagoort, P., & Bastiaansen, M. C. (2009). Trial-by-trial coupling between EEG and BOLD identifies networks related to alpha and theta EEG power increases during working memory maintenance. Neuroimage, 44(3), 1224-1238.
33. Seabold, S., & Perktold, J. (2010, June). Statsmodels: Econometric and statistical modeling with python. In Proceedings of the 9th Python in Science Conference (Vol. 57, p. 61).
34. Schneider, W., & Shiffrin, R. M. (1977). Controlled and automatic human information processing: I. Detection, search, and attention. Psychological review, 84(1), 1.
35. Takeuchi, H., Taki, Y., Nouchi, R., Hashizume, H., Sekiguchi, A., Kotozaki, Y., ... & Kawashima, R. (2013). Effects of working memory training on functional connectivity and cerebral blood flow during rest. Cortex, 49(8), 2106-2125.
36. Thompson, J. E., Egger, S., Böhm, M., Haynes, A. M., Matthews, J., Rasiah, K., & Stricker, P. D. (2014). Superior quality of life and improved surgical margins are achievable with robotic radical prostatectomy after a long learning curve: a prospective single-surgeon study of 1552 consecutive cases. European urology, 65(3), 521-531.
37. Tucker, M. A., Morris, C. J., Morgan, A., Yang, J., Myers, S., Pierce, J. G., ... & Scheer, F. A. (2017). The relative impact of sleep and circadian drive on motor skill acquisition and memory consolidation. Sleep, 40(4), zsx036.
38. Van Dillen, L. F., Heslenfeld, D. J., & Koole, S. L. (2009). Tuning down the emotional brain: an fMRI study of the effects of cognitive load on the processing of affective images. Neuroimage, 45(4), 1212-1219.
39. Van Gerven, P. W., Paas, F., Van Merriënboer, J. J., & Schmidt, H. G. (2004). Memory load and the cognitive pupillary response in aging. Psychophysiology, 41(2), 167-174.
40. Van Gog, T., & Paas, F. (2008). Instructional efficiency: Revisiting the original construct in educational research. Educational psychologist, 43(1), 16-26.
41. Van Gog, T., & Scheiter, K. (2010). Eye tracking as a tool to study and enhance multimedia learning.
42. Van Merrienboer, J. J., & Sweller, J. (2005). Cognitive load theory and complex learning: Recent developments and future directions. Educational psychology review, 17(2), 147-177.
43. Walker, M. P., Brakefield, T., Morgan, A., Hobson, J. A., & Stickgold, R. (2002). Practice with sleep makes perfect: sleep-dependent motor skill learning. Neuron, 35(1), 205-211.
44. Wang, S., Gwizdka, J., & Chaovalitwongse, W. A. (2015). Using wireless EEG signals to assess memory workload in the n-back task. IEEE Transactions on Human-Machine Systems, 46(3), 424-435.
45. Ye, J., & Li, Q. (2005). A two-stage linear discriminant analysis via QR-decomposition. IEEE Transactions on Pattern Analysis and Machine Intelligence, 27(6), 929-941.
46. Yiannakopoulou, E., Nikiteas, N., Perrea, D., & Tsigris, C. (2015). Virtual reality simulators and training in laparoscopic surgery. International Journal of Surgery, 13, 60-64.
47. Zarjam, P., Epps, J., & Chen, F. (2011, August). Characterizing working memory load using EEG delta activity. In 2011 19th European Signal Processing Conference (pp. 1554-1558). IEEE.